\documentclass{aa}
\usepackage{graphicx,amssymb,amsmath,natbib,multirow}

\usepackage{graphicx}
\usepackage{txfonts}

\newcommand{\hi}{\mbox{H{\sc i}}}
\newcommand{\kms}{km~s$^{-1}$}

\newcommand{\st}{$\sigma_\phi$}
\newcommand{\fract}{$\sigma_\phi/\sigma_R$}

\newcommand{\sr}{$\sigma_R$}
\newcommand{\sz}{$\sigma_z$}
\newcommand{\spp}{$\sigma_{\phi}$}
\newcommand{\splan}{$\sigma_P$}
\newcommand{\sth}{$\sigma_T$}

\newcommand{\slos}{$\sigma_{\rm los}$}
\newcommand{\slossq}{$\sigma^2_{\rm los}$}
\newcommand{\slosmean}{$\langle \sigma_{\rm los} \rangle_\phi$}

\newcommand{\betaea}{$\beta_{\rm EA}$}
\newcommand{\betat}{$\beta_\phi$}

\bibpunct{(}{)}{;}{a}{}{,} 
\usepackage{rotating}
\usepackage{txfonts}
 
\begin{document}

\title{Anisotropy of random motions of gas in Messier 33}

\titlerunning{Anisotropy of random motions of gas in Messier 33}
\author{Laurent Chemin\inst{1} \and Jonathan Braine\inst{2} \and Fran\c{c}oise Combes\inst{3} \and Zacharie S. Kam\inst{4} \and Claude Carignan\inst{4,5}}
\institute{Centro de Astronom\'ia, Universidad de Antofagasta, Avda. U. de Antofagasta 02800, Antofagasta, Chile\\ \email{laurent.chemin@uantof.cl, astro.chemin@gmail.com} 
          \and Laboratoire d'Astrophysique de Bordeaux, Univ. Bordeaux, CNRS, B18N, all\'ee Geoffroy Saint-Hilaire, 33615 Pessac, France 
          \and Observatoire de Paris, LERMA, College de France, CNRS, PSL Univ., Sorbonne Univ., UPMC, Paris, France
          \and Laboratoire de Physique et de Chimie de l'Environnement, Observatoire d’Astrophysique de l'Universit\'e Ouaga I Pr Joseph Ki-Zerbo (ODAUO), 03 BP 7021, Ouaga 03, Burkina Faso
          \and Department of Astronomy, University of Cape Town, Private Bag X3, Rondebosch 7701, South Africa}

   \date{Accepted 2020 May 19}

 \abstract
 { The ellipsoid of random motions of the gaseous medium in galactic disks is often considered isotropic, as appropriate if the gas is highly collisional. 
 However, the collisional or collisionless behavior of the gas is a subject of debate. 
 If the gas is clumpy with a low collision rate, then the often observed asymmetries in the gas velocity dispersion could be hints of anisotropic motions in a gaseous collisionless medium.}
 {We study the properties of anisotropic and axisymmetric velocity ellipsoids from maps of the gas velocity dispersion in nearby galaxies.
  This data allow us to measure the azimuthal-to-radial axis ratio of gas velocity ellipsoids, 
 which is a useful tool to study the structure of gaseous orbits in the disk. 
 We also present the first estimates of perturbations in gas velocity dispersion maps by applying an alternative model that considers isotropic and asymmetric random motions. }
{High-quality velocity dispersion maps of the atomic  medium at various angular resolutions of the nearby spiral galaxy Messier 
33, are used to test the anisotropic and isotropic velocity models.  The velocity dispersions of hundreds of individual molecular clouds are also analyzed.}
{ The \hi\  velocity dispersion of M33 is systematically larger along the minor axis, and lower along the major axis. 
Isotropy is only possible if asymmetric motions are considered. Fourier transforms of the \hi\ velocity dispersions reveal 
a bisymmetric mode which is  mostly stronger than other asymmetric motions and aligned with the minor axis of the galaxy. 
  Within the anisotropic and axisymmetric velocity model,  the stronger bisymmetry is explained by a radial  
 component  that is larger than the azimuthal component of the ellipsoid of random motions, thus by gaseous orbits that are dominantly radial. 
 The azimuthal anisotropy parameter is  not strongly dependent on the choice of the vertical dispersion.  The velocity anisotropy parameter
 of the molecular clouds is observed highly scattered. }
{Perturbations such as  \hi\  spiral-like arms  could be at the origin of the gas velocity anisotropy in M33. 
Further work  is necessary to assess whether anisotropic velocity ellispsoids can also be invoked to explain the asymmetric gas random motions of other galaxies. }
 

   \maketitle

\section{Introduction}

\label{sec:intro}
  What is the structure of orbits in the interstellar medium of disk galaxies?  What is the shape of the gaseous velocity dispersion ellipsoid in the mid-plane of galactic disks?

A quick look at observations of nearby galaxies is helpful to guess that orbits should not be perfectly circular, owing to 
the presence of large-scale   perturbations like bars, spiral arms, warps, or disk lopsidedness. 
Rotational motions are accompanied by radial and streaming motions, and orbits become asymmetric \citep{kal73, vis80}. 
In a galactic disk,    the structure of  orbits can be studied by means of the shape of the velocity dispersion ellipsoid \sr, \spp, and \sz, which are respectively the radial, tangential, and vertical components of random motions in cylindrical coordinates. 
The ellipsoid is characterized by two axis ratios, one of  which is the azimuthal-to-radial  ratio
that traces the degree of anisotropy in the mid-plane, as characterized by the azimuthal anisotropy parameter, $\beta_\phi = 1 - (\sigma_\phi/\sigma_R)^2$. 
Orbits that are biased tangentially have $\beta_\phi < 0$, with $\beta_\phi \rightarrow - \infty$ for circular orbits,  
whereas those biased radially have $0< \beta_\phi < 1$ \citep{bin08}. This parameter is appropriate to explain the orbital structure of a kinematic tracer that is collisionless (stars). 
We would expect to observe such values  within gaseous disks of galaxies, if interstellar gas behaved partly like a collisionless medium, as shown in numerical 
models \citep{bot03,age09}.
    
Interstellar gas is a medium in which collisions and shocks coming from all directions are thought to completely ``isotropize'' the motions.  
Assuming a velocity ellipsoid that is not tilted,  the line-of-sight dispersion can be written 
\begin{equation}
 \sigma_{\rm los} = \big ( (\sigma^2_R \sin^2\phi + \sigma^2_\phi \cos^2\phi) \sin^2 i +  (\sigma_z \cos i)^2 + \sigma^2_T + \sigma^2_{\rm ins} \big)^{1/2}  
 \label{eq:sigmalos}
,\end{equation}
  where $i$ is the disk inclination, $\phi$ the azimuthal angle in the disk plane, 
  \sth\ the thermal (isotropic) component, and $\sigma_{\rm ins}$ the instrumental broadening. {Isotropy 
  greatly simplifies   the modeling of the observed dispersion as it implies $\sigma_R =\sigma_\phi=\sigma_z=\sigma_{\rm iso}$, reducing Eq.~\ref{eq:sigmalos} to 
   \begin{equation}
    \sigma_{\rm los} = (\sigma_{\rm iso}^2 + \sigma^2_T + \sigma^2_{\rm ins})^{1/2} \, .
    \label{eq:sigmalosiso}
   \end{equation}  
 Important values for galactic dynamics like the gas radial pressure support  depending on \sr\ \citep{dal10,oh15} or the vertical equilibrium  depending on \sz\ \citep{com97, koy09,dms1,dms6}
 can then be easily estimated  directly from observations, without needing to deproject \slos.  } 
 
 \begin{figure*}
   \centering
  \includegraphics[width=18cm]{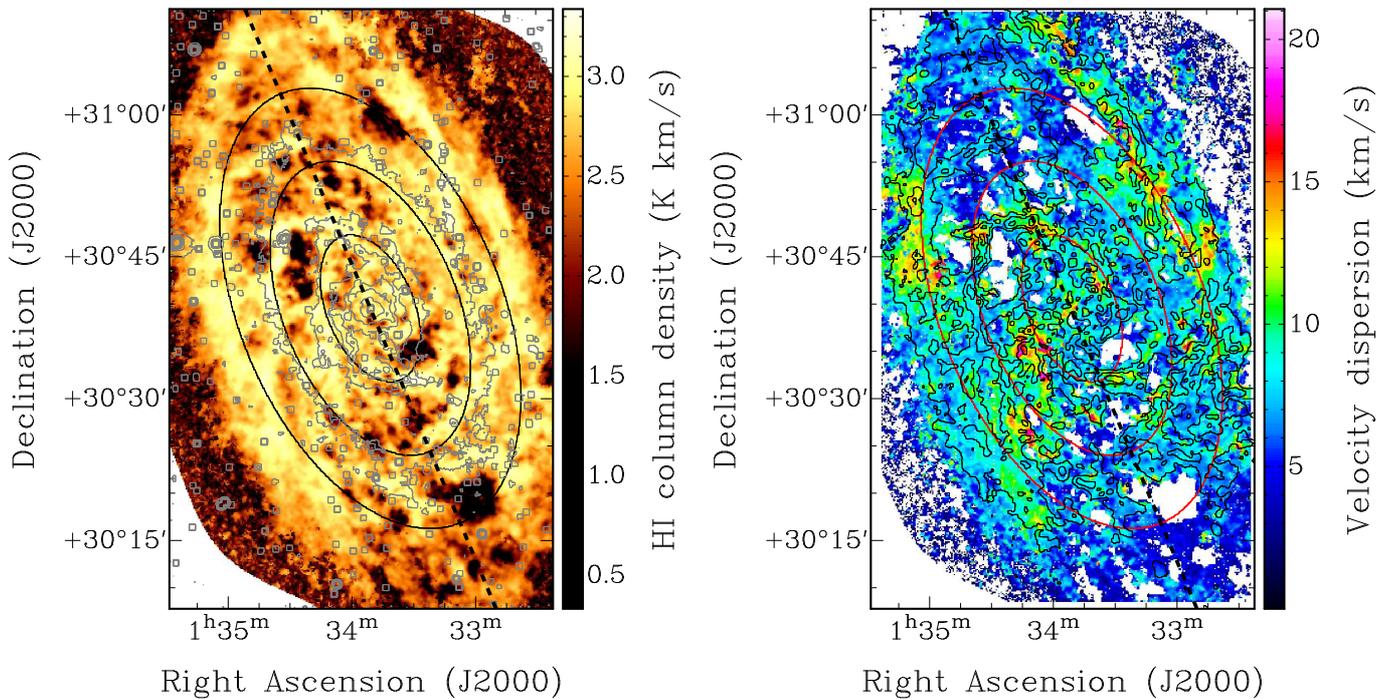}
  \caption{Observational data of Messier 33. Left panel:  \hi\ column density map (VLA, 100 pc resolution, logarithmic stretch)  
   with Spitzer/IRAC $3.6\, \rm \mu m$ stellar distribution overlaid (gray contours, showing the 0.1, 0.3, 0.5, 0.7, 0.9, 1.4, and 2.5 MJy/sr levels).
  Right panel: Observed \hi\ velocity dispersion (VLA, 100 pc resolution, not corrected for the instrumental dispersion), with \hi\ column densities  contours  ($1,2,3\times 10^{21}$ cm$^{-2}$).
  The dashed line represents the location of the major axis of the inner unwarped disk, with position angle of 22.5\degr. Concentric ellipses show the projected locations of $R=2, 4, 6$ kpc.}
  \label{fig:imam33}
 \end{figure*}
   
   The analysis of \slos\ is less straightforward in the hypothesis that part of the gas behaves as a collisionless medium. Equation~\ref{eq:sigmalos} has to be  
       modeled or inverted in order to fit or deproject (respectively) the data,  and constrain the shape of the velocity ellipsoid. Moreover, Eq.~\ref{eq:sigmalos}   
   predicts that \slos\ is a function  of the azimuthal angle owing to the projection of the radial and tangential components into \slos. 
   In an axisymmetric approach of this kind, an anisotropic dispersion ellipsoid  has  a significant implication on \slos: it has to be asymmetric, greater near  
   the minor (major) axis for $\sigma_\phi<\sigma_R$ ($\sigma_\phi>\sigma_R$, respectively).
   To our knowledge, such strong prediction has never been tested. 
   Many observed velocity dispersion fields exhibit asymmetric \slos\ dependent on the azimuth  \citep[Fig.~\ref{fig:imam33}, and][]{wal08}, 
   and some of them are reminiscent of this particular anisotropic signature.

  The two competing collisional and  collisionless models for the interstellar medium thus predict different shapes for the velocity dispersion ellipsoid, respectively   isotropic and anisotropic, 
   that are important to study from an observational viewpoint. 
   At the same time, the traditional assumption of axisymmetric motions under the isotropy argument seems not suitable to asymmetric observations as it is independent of azimuth, by construction. 
      It is   therefore important to evaluate the degree of asymmetry in velocity dispersion fields as well.

      We want to address these problems in this work. Our long-term objective is to determine observationally whether collisional or collisionless models are appropriate to explain 
      the structure of gaseous velocity dispersion fields. We  want 
      to study the asymmetries in velocity dispersion maps for the atomic and/or molecular gas in nearby galaxies, and estimate the azimuthal velocity
      anisotropy parameter  that is needed to reproduce asymmetric observations.  We start this study with the Local Group spiral Messier 33 (M33). Its proximity ($D=840$ kpc) has made 
   it possible to obtain very high-quality data in the \hi\ and CO lines, tracing the atomic and molecular gas components  (Sect.~\ref{sec:data}). 
   Velocity dispersion maps  are used to quantify asymmetries assuming  an isotropic velocity model (Sect.~\ref{sec:isomodel}), 
   and estimate  \betat\  and the corresponding structure of gas orbits for the atomic and molecular gas in  M33 {assuming an 
   anistropic velocity model} (Sect.~\ref{sec:anisomodel}). 
   A short discussion on the two competing models and comparisons with numerical models and  the velocity anisotropy expected in the framework of the epicycle theory of collisionless 
   orbits is also presented (Sect.~\ref{sec:discussion}).

 \section{Atomic and molecular gas data of M33} 
\label{sec:data}

The \hi\ data  of M33 come from three studies.  Our main \hi\ reference is the  25\arcsec\ resolution (100 pc) Very Large Array (VLA) data from \citet{gra10}.
We also use other recent VLA observations  at 18\arcsec\ resolution (70 pc) by \citet{koc18}.
Both data sets are very low noise, which is essential to calculate reliable second-moment maps (velocity width).
Even though   from the same telescope (the VLA), the data are from completely independent observations.  
We also use the low-resolution Dominion Radio Astrophysical Observatory (DRAO) data set by \citet[][2\arcmin, 490 pc]{kam17} as a further check. Both  are useful to estimate any link between resolution and the anisotropies presented here. 
The pixel size of the VLA velocity dispersion maps used in this study is 8\arcsec, and that of the DRAO map is 21.8\arcsec.

 Velocity widths via calculation of the second moment of a spectrum are extremely subject to noise, particularly the noise in channels far from the line center.
The velocity dispersion (second moment of a spectrum) is  $\sigma_{\rm los}$ by $\sigma_{\rm los}^2 = \frac{ \int_{v_{min}}^{v_{max}} T (v-v_{cen})^2 dv } { \int_{v_{min}}^{v_{max}} T dv }$, 
where $v_{cen}$ is the central (intensity-weighted mean) velocity.  
The $(v-v_{cen})^2$ term makes it critical to exclude noise beyond the line profile.  
Thus, we choose the low-noise \hi\ data at 100 pc resolution presented in \citet{gra10} to define the windows over which to calculate the moments. Over 90\% of the surface of M33 reaches a line temperature of 10~K 
in \hi, corresponding to approximately four times the noise level in the data.  We do not calculate the velocity dispersion where the line temperature does not reach 10~K 
because the uncertainty becomes too great, both on  $v_{cen}$  and thus on the dispersion. The 10~K cut applies only for the 100 pc resolution data.

The integration window is defined by taking the maximum and then descending to either side (higher and lower velocities) 
until a channel goes below zero.  In less than 1\% of the disk,  a double-peaked profile can be observed,  but at 100 pc resolution there are 
no pixels for which the line temperature goes below zero  between the two peaks. 
A mask is  created  for each position, containing a value for  $v_{min}$ and  $v_{max}$, and a flag indicating whether the line temperature reaches 10~K. 
This is the mask used in all the second-moment calculations of the \hi\ gas. 
We note that the velocity dispersions for the 70 pc and 490 pc resolution data are  $\sim 1$ \kms\ larger than those measured 
with the 100 pc resolution VLA data. As single-dish data have been merged to the 70 pc VLA and 490 pc DRAO interferometric data, but not to the 100 pc VLA data, 
combined data are more sensitive to larger scale structures that slightly widen the \hi\ profile wings. This small difference has no impact on the results. 
Figure~\ref{fig:imam33} shows examples of column density and velocity dispersion fields for the atomic gas (100 pc resolution VLA data).
 
The \hi\ disk of M33 is known to locally exhibit emission with anomalous velocities \citep{kam17,koc18}. This emission, which has typical characteristics of 
 an extraplanar \hi\ layer lagging the rotation of the disk \citep[e.g., in the galaxy NGC 2403;][]{fra01,fra02} with forbidden velocity gas 
 and high-velocity components \citep[see Sect. 3.2. of][]{kam17}, could participate in broadening  the \hi\ profiles. Quantifying the effect of such gas on the \hi\ 
 velocity dispersion models has not been attempted, however.

The CO observations are at 12\arcsec\ resolution (50 pc) and were made at the 30m telescope of the Institut de Radioastronomie Millim\'etrique (IRAM), as described in detail in \citet{dru14}.  
The velocity dispersions  for the molecular gas are not strictly derived   the same way as for \hi.
The molecular gas of M33 is   organized into clouds. Although there is emission which appears  diffuse, we do not know whether it is truly diffuse molecular gas or small clouds producing a weak but extended CO signal.  
Either way, the signal-to-noise ratio for the ``diffuse'' component is too low to try to calculate a velocity dispersion map from the IRAM datacube. 
Therefore, we use the positions and  velocity dispersions of the 566 clouds presented in \citet{cor17} and \citet{bra18} rather than a continuous map of the velocity dispersion. The decomposition
into clouds presented in these articles was made using CPROPS \citep{ros06}.

The instrumental dispersions of the \hi\ and CO observations is $\sigma_{\rm ins} \lesssim 1$ \kms.  In the kinematic modeling, the isotropic thermal component \sth\
 was chosen to be  constant with radius. For the atomic gas, fits were performed assuming two extreme cases: \hi\ seen as a cold  or warm   neutral medium \citep[CNM and WNM, respectively with $T \sim 100$ K and $T \sim 5000$ K,][]{dra11}. The CNM has 
 a thermal dispersion of $\sim 0.9$ \kms, which is comparable to the instrumental dispersion and significantly smaller than the observed \hi\ \slos. For simplicity, we thus set $\sigma_T =0$ \kms\ for the CNM case. 
 The WNM has a thermal dispersion of $6.4$ \kms, thus larger than the instrumental dispersion, and  comparable to \slos\ in many disk regions. For simplicity we thus set $\sigma_T = 6$ \kms\ for the WNM case. 
 Given that the majority of fits at $\sigma_T =6$ \kms\ failed (see below)
  most of results shown hereafter are those obtained for the CNM case, unless specified. The fits to the molecular gas dispersion were made assuming only a cold medium ($\sigma_T = 0$ \kms).

 This work focuses only on the radial range $R \le 7.5$ kpc that is not affected by the warping of the gaseous disk \citep{cor97,cor14,kam17}.
 The projection of the dispersion model of Eq.~\ref{eq:sigmalos} along the line of sight thus assumed 
 a fixed inclination of $56\degr$, and a fixed orientation of the  major axis shown as a dashed line in Fig.~\ref{fig:imam33}, with a position angle of $22.5\degr$ 
 ($\phi =0$ is  chosen aligned with the semi-major axis of the approaching half to the northeast,   
 and increases in the counterclockwise direction).  
 The fits were performed  at radii starting from $R=20\arcsec, 25\arcsec$, and $120\arcsec$  for the 70 pc, 100 pc, and 490 pc resolution data, respectively, with a radial bin width of $20\arcsec, 25\arcsec$ and $120\arcsec$. 
 The fits were performed  at radii starting from $R=60\arcsec$ for the molecular gas, with a radial bin width of 120\arcsec. 
 Fits were performed within the radial range  $R \la 1500\arcsec$   for the molecular gas, using 553 from the initial 566 clouds. We also note  that the 
 sampling is smaller than for the VLA data, although the CO data were obtained at higher angular resolution. 
 The smaller sampling for the discrete molecular gas clouds is necessary to have a number of degrees of freedom large enough to yield successful least-squares fits.
  For both components we thus excluded rings that had fewer than 10 points.
  
    \begin{figure*}
      \centering
  \includegraphics[width=\textwidth]{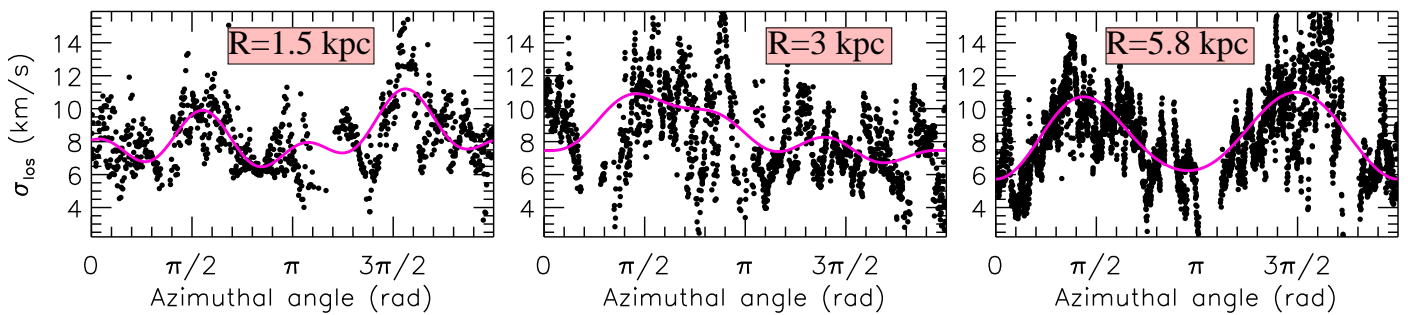}
  \caption{{Azimuth-velocity dispersion diagrams of atomic gas at selected radii in M33. 
   Each point corresponds to an individual measurement, i.e., the dispersion of each pixel within the map. 
   Solid lines represent the results of a Fast Fourier Transform model of the \hi\ dispersion map (Sect.~\ref{sec:isomodel}, assuming a cold neutral medium case).
   The dispersions are from the VLA 100 pc resolution data.}  }
    \label{fig:plotazimisoasym}
 \end{figure*} 
 
The telescope beams cover a larger area near the minor axis than on the major axis so the beam-smearing 
could induce a slightly higher velocity dispersion along the minor axis, and thus an apparent (but false) anisotropy \citep{che18}.
 The estimation of the beam-smearing effect was performed using a high-resolution model of gas intensity and velocity of M33 smoothed to the native resolution of observations (100 and 490 pc), 
 following prescriptions given in \citet{epi10}. In particular, we used the rotation curve model given in \citet{koc18}, and assumed a constant surface density  \citep{kam17} 
 and constant velocity dispersion (\slos$= 8$ \kms, Figs.~\ref{fig:testisotropy} and ~\ref{fig:sigphisigrhi}) within $R \le 7.5$ kpc. 
 We found that the contamination of the VLA dispersion map by the smearing effect is completely negligible along the principal axes ($<<1$ \kms). 
  The effect is stronger at a resolution of 490 pc, but the agreement (Fig.~\ref{fig:anishi}) between  the resolutions shows that the beam-smearing is not an issue. 
    For simplicity we assumed that artificial changes in dispersion with azimuth  caused by other mechanisms, like possible finite disk thickness \citep{bac19}, are negligible.
   
  Finally, we add that the models presented in Sect.~\ref{sec:isomodel} and Sect.~\ref{sec:anisomodel} 
   can be applied to linear  or squared   velocity dispersions. In the squared velocity case, Eqs.~\ref{eq:sigmalos} and~\ref{eq:sigmalosiso} resume to simple additions which 
   could make the analysis easier at first glance.  We thus performed the modeling of both \slos\ and \slossq\ and found no significant difference  
   between results from the two approaches. Hereafter, the presented results are those obtained for the modeling of \slos.   
 
\section{Evidence of asymmetric line-of-sight \hi\ velocity dispersions in M33}
 \label{sec:isomodel} 
 
  Asymmetries are obvious   in the \hi\ velocity dispersion map of M33 (Figs.~\ref{fig:imam33} and~\ref{fig:plotazimisoasym}).  
  The variations in \slos\ with the azimuthal angle occur at all angular scales.  There is for example a clear hint of a large angular scale variation for $R> 4$ kpc, 
  with \slos\ lower towards the major axis of the disk and larger towards the minor axis. In that region the difference between individual pixel values \slos\ and 
  the azimuthally averaged dispersion \slosmean\ at the same location is easily $>25\%$ of \slos, and can sometimes reach up to $60\%$. 
  A mean axisymmetric and isotropic dispersion is thus clearly ruled out to explain the observed  variations occurring on larger angular scales (see also Sect.~\ref{sec:anisomodel}).  
  The  anisotropic and isotropic   models presented below attempt to capture such large angular scale variations, not those occurring on angular
  scales smaller than a few  degrees (like the pixel-to-pixel variations). 
  
   To investigate the  properties of the asymmetric random motions of the atomic gas, we expanded the isotropic random motions $\sigma_{\rm iso}$ of Eq.~\ref{eq:sigmalosiso} with Fourier coefficients, 
   following $\sigma_{\rm iso, asym} = \sum_k \sigma_k \cos k(\phi -\phi_k)$, where $k$ is an integer.     
   The isotropic axisymmetric component is  $\sigma_0$ and  $\sigma_k$ and $\phi_k$ are the amplitude and phase of the asymmetric modes in the velocity dispersion map. 
   Both least-squares (LSQ) fits and discrete fast Fourier transforms (FFTs)  were performed at the radial rings defined in Sect.~\ref{sec:data} to  cross-validate the analysis.
    The minor differences between fits and FFTs are that fits provide formal errors for the phases and amplitudes,   
   and are more time consuming than FFTs. To avoid possible degeneracies, divergence,  and increased computational time during minimizations, the maximum   order that was fit is $k=4$  (harmonics of $k=2$). This maximum order
   is much smaller than the last order yielded by FFTs, which is the number of 
   points in the azimuthal dimension at a given radius.  The instrumental and thermal contributions were subtracted from the original maps before derivations of the kinematic harmonics.
  
\begin{figure}
      \centering
  \includegraphics[width=8cm]{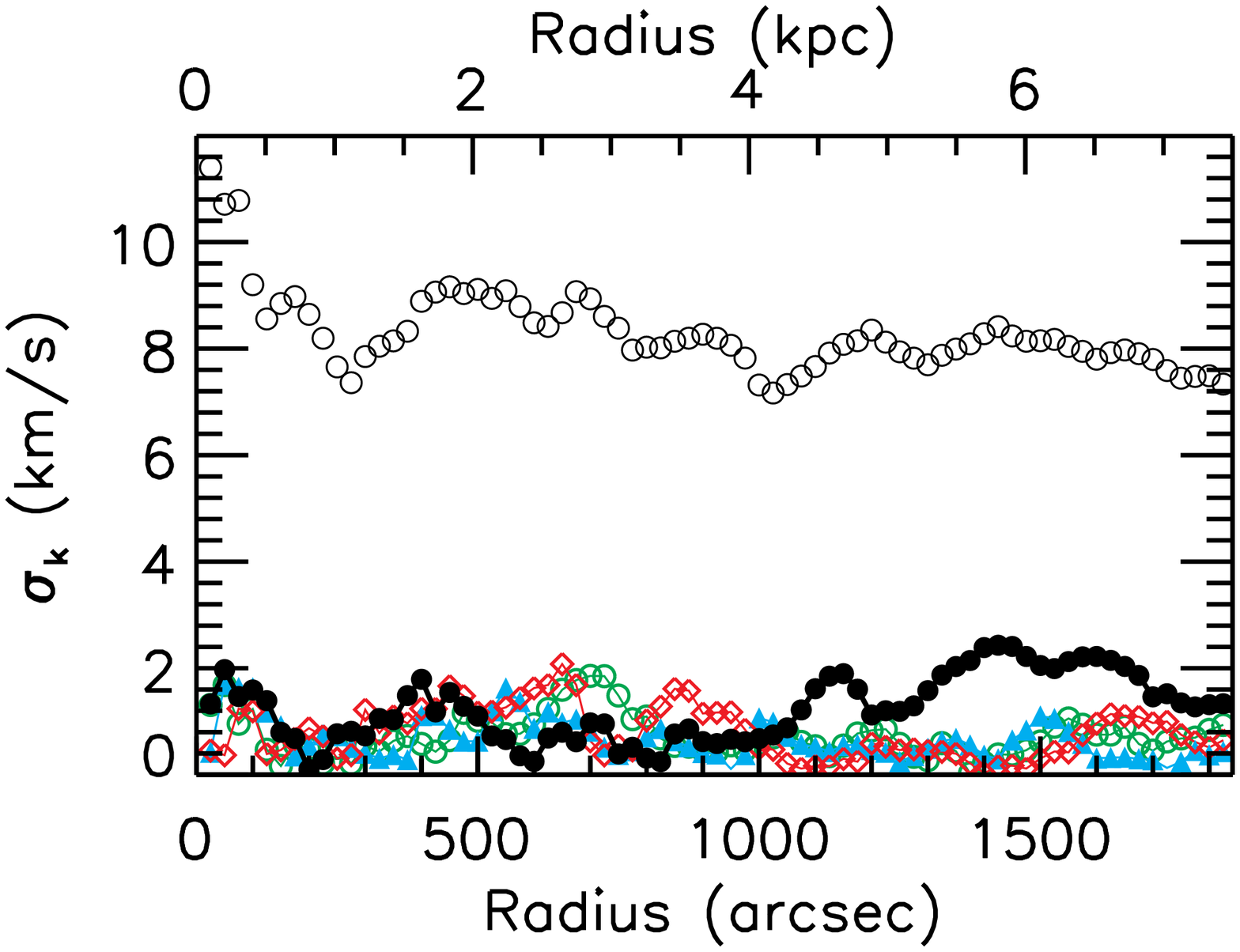}
  \includegraphics[width=8cm]{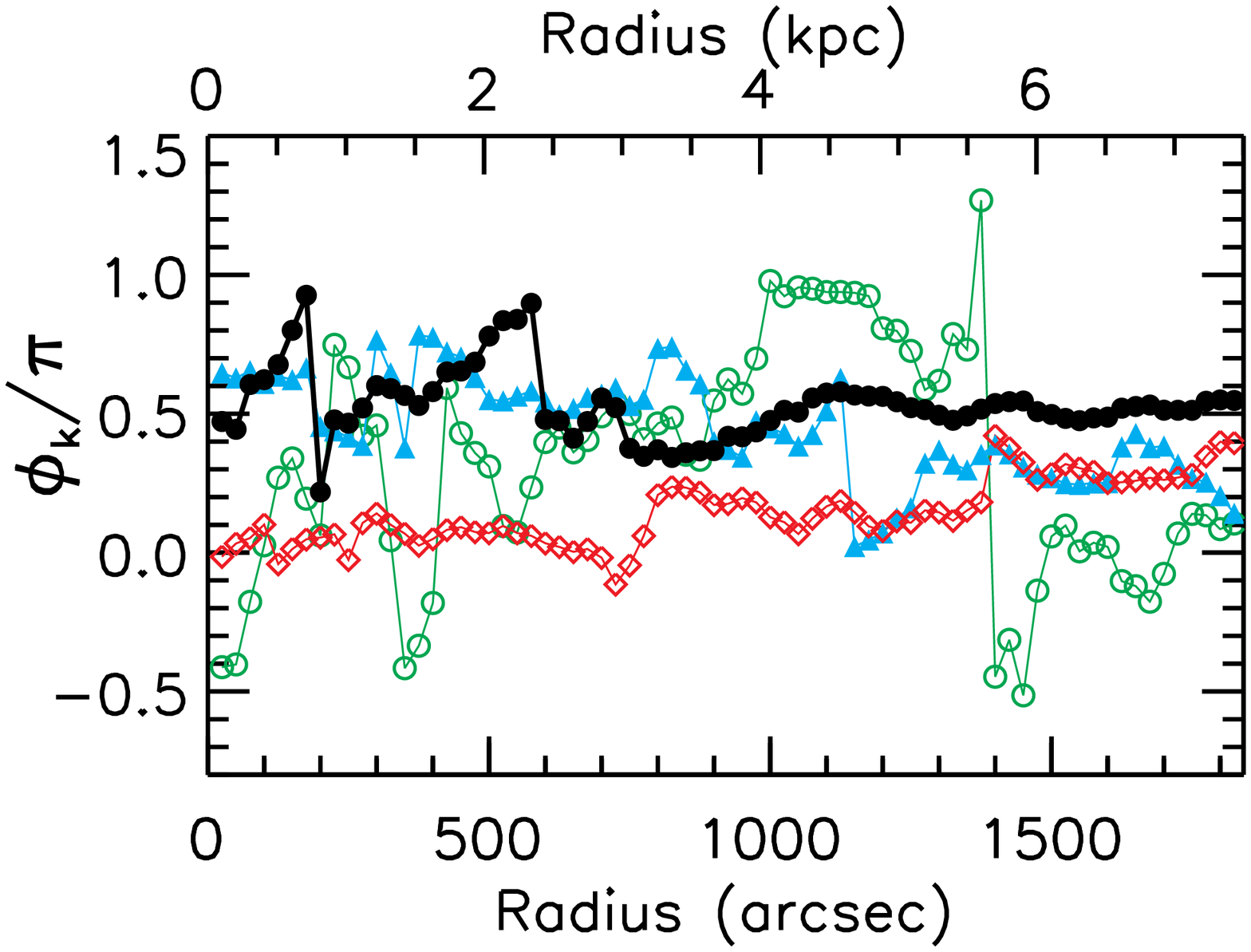}
  \includegraphics[width=8cm]{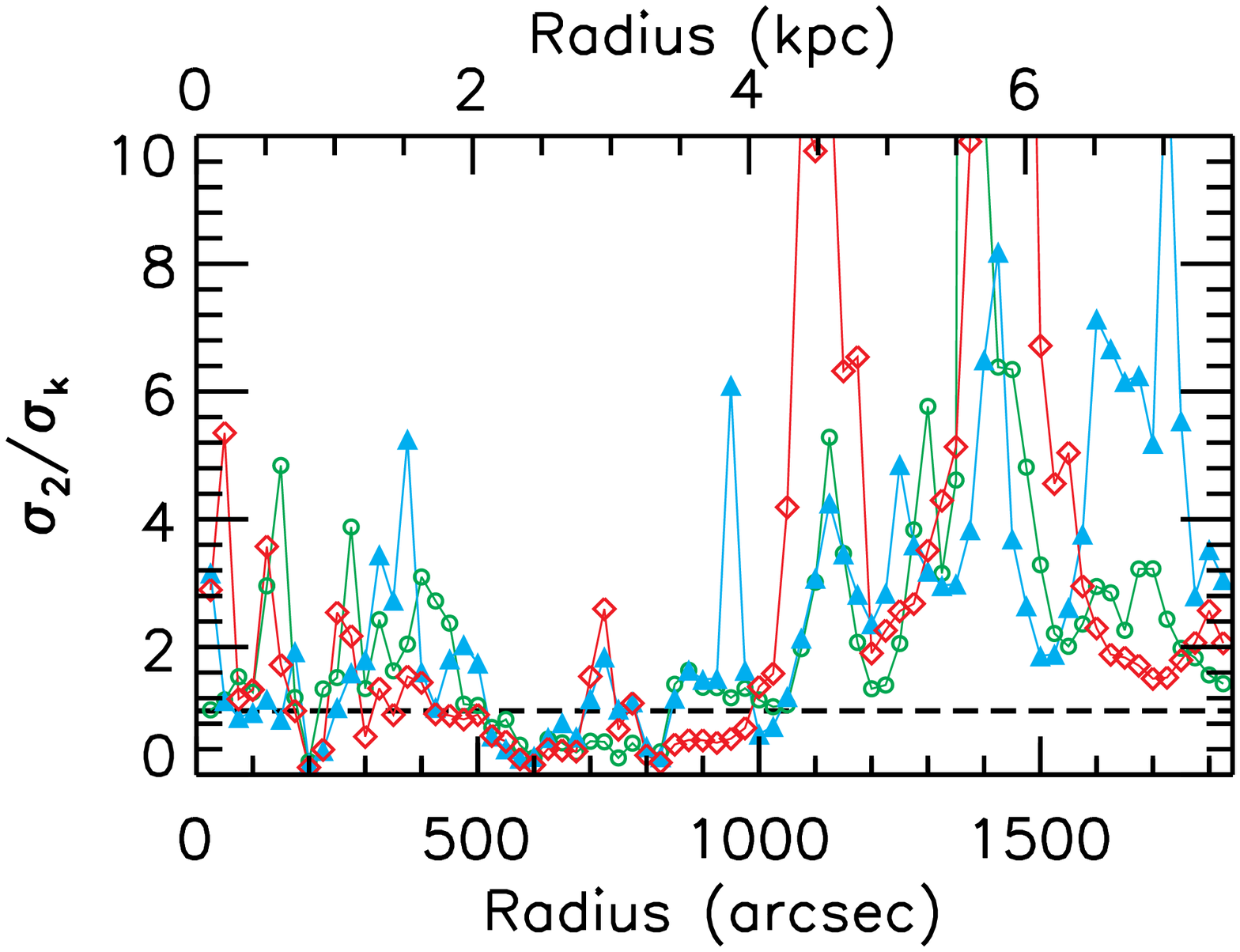}
  \caption{Results of the discrete FFT of the \hi\ velocity dispersion map of M33 (100 pc resolution, in the CNM case). 
  The amplitudes (upper panel) and phases (middle panel) of the Fourier modes are 
    shown as black open circles ($k=0$), green open circles ($k=1$, joined by a solid line), filled circles ($k=2$), blue upward filled triangles ($k=3$), and red open diamonds ($k=4$).  
    The reference $\phi=0$ is chosen along the semi-major axis of the NE approaching half of M33. The bottom panel shows $\sigma_2/\sigma_k$, 
    using the same colors and symbols as above for the $k=1, 3, 4$ modes. In the bottom panel, the range of ratios is chosen up to 10 for clarity, but the 
    total range is 78 for $\sigma_2/\sigma_1$ and $\sigma_2/\sigma_4$, and 12 for $\sigma_2/\sigma_3$.}
    \label{fig:asymisofft}
 \end{figure}

   Figure~\ref{fig:asymisofft} shows the results of the FFT derivation for the 100 pc resolution VLA observation in the CNM case, with the amplitudes (top panel) and
   phases (middle panel) of the first four Fourier modes. The comparison of the bisymmetric amplitude with the first-, third-, and fourth-order terms is shown in the bottom panel of 
   Fig.~\ref{fig:asymisofft}. Results of the LSQ fits are not shown because they are extremely similar to the FFTs. 
   
   The amplitude of the axisymmetric term is unsurprisingly  similar to the azimuthally averaged velocity dispersion.
   The other modes are weak, with median amplitudes of $\sigma_1 = 0.6$ \kms, $\sigma_2 = 1.2$ \kms, $\sigma_3 = 0.6$ \kms, and $\sigma_4 = 0.7$ \kms.
   The bisymmetry is the most interesting mode. 
   It shows the largest variation, which can reach 25\%\ of the axisymmetric value at around $R=1500\arcsec$ (6 kpc).  
   It is stronger than  $\sigma_1,\sigma_3$, and $\sigma_4$  in  $\sim$80\%, 80\%, and 65\%\ of the cases, respectively. 
   In the inner region ($R < 1000\arcsec$, or 4 kpc), these ratios fall to 65\% for the $k=1$ and $k=3$ amplitudes, and to 40\% for $k=4$. 
   We can thus define the locations where the bisymmetry is not stronger than other asymmetries as the locations where $\sigma_2$ is  weaker than at least two other modes, that is   
   $188 \arcsec \le R \le 237\arcsec$, $513\arcsec \le R \le 687\arcsec$, and $737\arcsec \le R \le 837\arcsec$. These regions are highlighted in subsequent figures.
   In the outer region ($R > 1000\arcsec$) it significantly dominates  the  other modes, by factors of 5, 4, and 7 on average (respectively to the $k=1,3,4$ modes), 
   reaching up to 78 times the $k=1$ and $k=4$ terms, and 12 times the $k=3$ term. 
   
     The phases of the asymmetries contain a wealth of information as well. They show extended angular ranges where they
  remain closer or aligned with one of the principal axes of the M33 disk,  modulo half or one period of each mode. 
  In other words, $\phi_1$  is consistent with the position of the minor and major axes ($2.4 < R < 3.4$ kpc, $4 < R < 5$ kpc, $R> 6$ kpc) with some variation elsewhere, 
   $\phi_2$ is mostly aligned with the minor axis, except for $R = 0.7$ and $1.5-2.5$ kpc,  $\phi_3$ remains close to $2\pi/3$ in the inner disk half and 
   to $\pi/3$ beyond $R=5.2$ kpc (the period and half-period of the $k=3$ mode, respectively),  while $\phi_4$ shows three steps, one close to the major axis for $R<3$ kpc 
   ($\phi_4 \sim 0$), one with the half-period of the $k=4$ mode for $3 < R < 5.5$ kpc ($\phi_4 \sim \pi/4$) and another close to $\pi/3$ beyond 5.5 kpc.
    In the azimuth-dispersion diagrams drawn at selected radial rings of 0.2 kpc in width (Fig.~\ref{fig:plotazimisoasym}), 
    the solid lines show the results of the FFT models, highlighting the effects of the
   bisymmetry (all panels), the $k=1$ mode ($R=3$ kpc, middle panel), and the $k=4$ mode ($R=1.5$ kpc, left panel). 
        
  Finally, comparable results have been obtained with the 70 pc and 490 pc data.
  We also performed the same analysis in the WNM case for the 100 pc resolution data. 
  The phases are comparable to the CNM case, but on average  $\sigma_0$  drops by $25\%$, while the median 
  $\sigma_1$, $\sigma_2$, $\sigma_3$, and $\sigma_4$ rises respectively by a factor of 1.2, 1.1, 1.4, and 1.6, though it remains small. The $k=2$ mode thus still dominates 
  the asymmetric random motions in the WNM case.

 \section{Velocity anisotropy of gas in M33}
 \label{sec:anisomodel}
 
 We now consider that gas can behave like a collisionless medium.  
The {axisymmetric and} anisotropic velocity model of Eq.~\ref{eq:sigmalos} can  be recast in
\begin{equation}
 \begin{split}
 \sigma_{\rm los} = \big( 0.5(\sigma^2_\phi-\sigma^2_R)\sin^2 i \, \cos 2\phi +  0.5(\sigma^2_\phi+\sigma^2_R)\sin^2 i + \\ (\sigma_z \cos i)^2 + \sigma^2_T + \sigma^2_{\rm ins} \big)^{1/2}
 \label{eq:sigmalos2}
 \end{split}
.\end{equation}The expression with $\cos 2\phi$ therefore implies that {such an} anisotropic {velocity} dispersion model looks like a bisymmetric 
mode that has a null phase, i.e., aligned with the major or minor axis (depending on the sign of $\sigma^2_\phi-\sigma^2_R$). 
  The significant  bisymmetric perturbation mostly aligned with the minor axis found in the \hi\ velocity dispersion map of M33 could thus be interpreted as a signature of velocity anisotropy, except maybe 
  for the clearly  identified regions of weaker $k=2$ mode (see Sect.~\ref{sec:isomodel}). The anisotropic velocity model
 makes it possible to constrain the two planar components  and study the structure of orbits of gas in M33 through the azimuthal anisotropy parameter $\beta_\phi$. 
 We also apply this model to observations of the molecular gas (Sect.~\ref{sec:resultmol}).
 
Non-linear LSQ fits of Eq.~\ref{eq:sigmalos} to \slos\ were performed using the program developed  in \citet{che18} 
for stellar disks. It fits  radial, tangential, and vertical components of the random motion ellipsoid to a vector of observed dispersions at a given radius.
A more developed model where the anisotropic dispersion components are asymmetric should also be considered to allow direct comparisons with 
Sect.~\ref{sec:isomodel}.   Yet such modeling is beyond the scope of the article.
We  refer to  \citet{che18} for more details of the minimization process and the validation of the methodology. 

 As \sr, \spp, and \sz\ are expected to be roughly comparable, degeneracies can occur 
 if  the model is fit blindly to the observations. To make the analysis possible, we chose to hold the vertical component constant 
 and let the radial and tangential dispersions vary freely. The value of  $\sigma_z$ can then be varied to measure its effect on the values of \sr\ and \spp.   
  Fits with \sz\ assumed constant as a function of $R$ were performed, as well as others where \sz\ varied with radius. In this latter case, it was
  fixed at the azimuthally averaged dispersion \slosmean\ (dashed lines in Figs.~\ref{fig:testisotropy},~\ref{fig:sigphisigrhi},  and ~\ref{fig:anismol}).  
 In the case when \sz\ was assumed constant,  iterations in the range $4 \leq \sigma_z \leq 10$ \kms\ were made for the atomic gas, sampled every 0.2 \kms. This broad range of values contains the 
 line-of-sight dispersion of $\sim 8$ \kms\ observed in nearly face-on \hi\ disks \citep{sho84,vdk84}, and which is a good proxy for \sz. 
  For the molecular gas, we explored  $2 \le \sigma_z \le 5$ \kms\  because the velocity dispersions are lower. 
   The resulting anisotropy parameters are not observed to be strongly dependent on the choice of \sz\ \citep[see also][]{che18}. 
   We define the quoted error on \sr\ and \spp\  at a given value of \sz\ and at a given radius 
   as the standard deviation of the posterior distribution of the parameters, and that for \betat\ as the propagation of the \sr\ and \spp\ uncertainties.

 \subsection{Results for the atomic neutral hydrogen}
  \label{sec:resultatom}
 
 \subsubsection{One-parameter model}
 \label{sec:resultisoatom}
\begin{figure}[t!]
      \centering
  \includegraphics[width=\columnwidth]{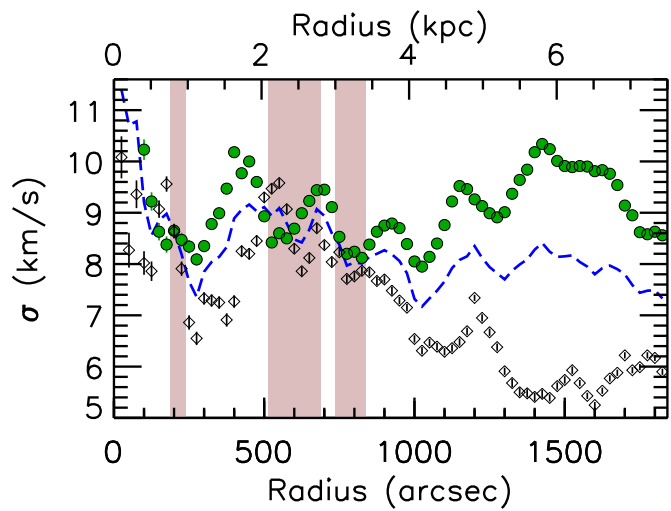}
  \caption{Results of the anisotropic and axisymmetric velocity model for the atomic gas in Messier 33 (100 pc resolution data). Results are those of the one-parameter model and show the profile of \sr\ (\st, respectively)
  as filled green circles (open diamonds),  obtained assuming $\sigma_\phi= \sigma_z = \langle \sigma_{\rm los} \rangle_\phi$  ($\sigma_R= \sigma_z = \langle \sigma_{\rm los} \rangle_\phi$).  
  A dashed blue line is for \slosmean\ (corrected from instrumental dispersion). Results obtained assuming a null thermal component. The shaded areas highlight the regions where the $k=2$ mode was found to be weaker than other 
  dispersion asymmetries (see Sect.~\ref{sec:isomodel}).}
    \label{fig:testisotropy}
 \end{figure}

 A first goal is to assess whether isotropy is present within the model of Eq.~\ref{eq:sigmalos}. In this context, we 
   tried to maximize the likelihood of finding isotropy by fixing \sr=\sz=\slosmean, with \spp\ as the only free parameter, and alternatively \spp=\sz=\slosmean\ with \sr\ as free parameter. We would indeed expect 
  $\sigma_\phi \simeq \langle \sigma_{\rm los} \rangle_\phi$ and $\sigma_R \simeq \langle \sigma_{\rm los} \rangle_\phi$, respectively, in those particular cases.
   The result is displayed in Fig.~\ref{fig:testisotropy} (for the CNM case). The locations where \sr\ or  \spp\ is closer to \slosmean\ are   $R\sim 1$ kpc, and  $2 \le R \lesssim 3.5$ kpc.  These radii match  
   closely those where the bisymmetry was found dominated by other asymmetries (Sect.~\ref{sec:isomodel}).
 The radial (azimuthal) component is observed to be larger (smaller, respectively) than   \slosmean\ in most disk regions.  Beyond 4 kpc, the difference with \slosmean\ is important,  up to 2 and 3 \kms\ for \sr\ and \spp\ (respectively), 
 which is significant compared to the formal errors of the fits ($<< 1$ \kms). Comparisons between the two isotropic fits show that the model at free \sr\ is significantly worse than the one at free \spp, particularly
 in the outer disk. Indeed the assumption \spp\ $=$ \slosmean\ automatically maximizes the tangential component, and in regions where the velocity dispersion map exhibits larger values near the minor axis, 
 the model has no other choice than finding an even larger radial component than \slosmean, hence yielding a  best-fit  \slos\ model that overestimates the bulk of the observed \slos.
 We also performed fits with $\sigma_T =6$ \kms\ (not shown), but they failed in more than 75\% of the rings. 
 Results for a   one-parameter case that most  closely approaches an axisymmetric and isotropic model thus confirm the results found in Sect.~\ref{sec:isomodel}: axisymmetry with isotropy cannot apply to
 the  \hi\ random velocity field of M33.

 \begin{figure}
 \centering
  \includegraphics[width=\columnwidth]{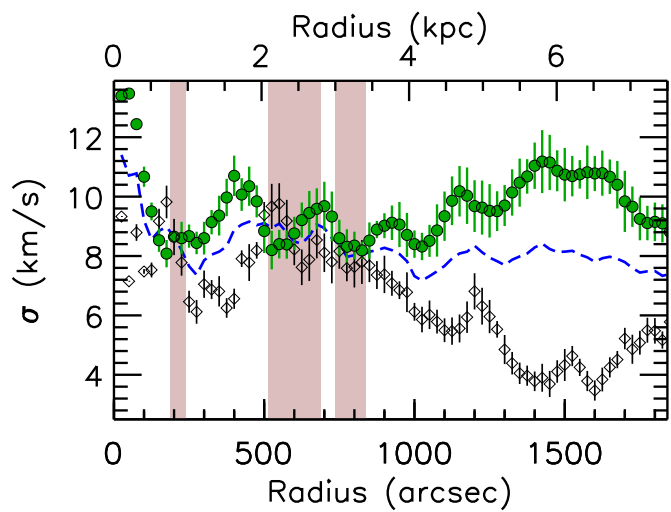}
  \caption{Same as Fig.~\ref{fig:testisotropy}, but for the two-parameter model with \sr\  and \st\ as free parameters.}
    \label{fig:sigphisigrhi}
 \end{figure} 
 
  \subsubsection{Two-parameter model}
  \label{sec:resultanisoatom}
     
  Figure~\ref{fig:sigphisigrhi} shows the   results found for $\sigma_R$ and $\sigma_\phi$ for the VLA 100 pc resolution map, assuming a null thermal component.  
  The results corresponding to the warm neutral medium case are briefly discussed below.
  Results for \sz=\slosmean\ only are shown for clarity.  
   Beyond $R=2$ kpc, the azimuthal dispersion decreases by $\sim 6$ \kms, while the radial dispersion tends to increase slightly, although bumps are observed. 
  In the inner kiloparsec, the overall variation of \sr\ reaches $\sim 5$ \kms. For the most realistic \sz\ values ($\sigma_z < 10$ \kms) the results yield  $0.5 \le \sigma_z/\sigma_R \le 0.9$.

  \begin{figure}[t!]
  \centering
  \includegraphics[width=\columnwidth]{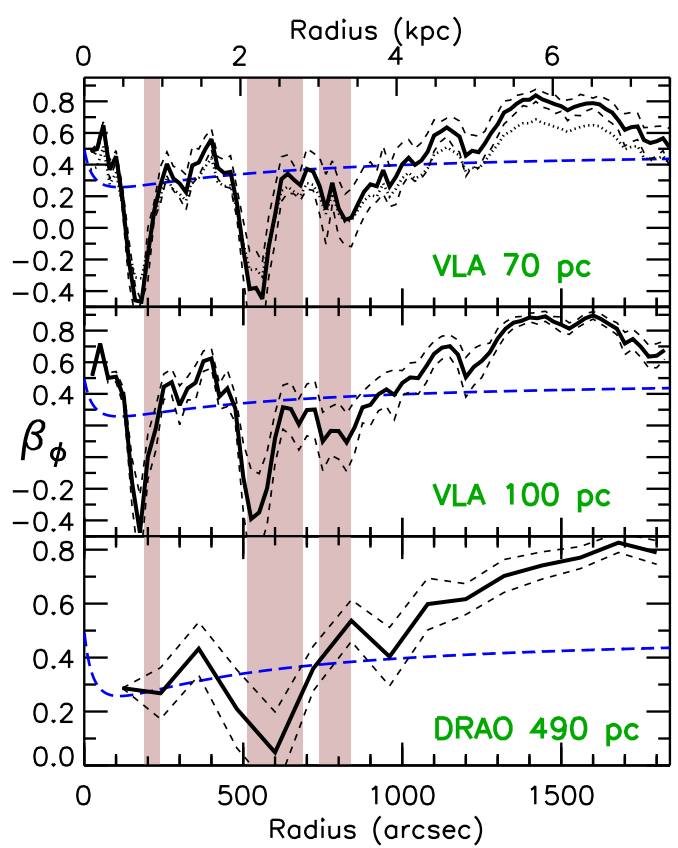} 
  \caption{Profiles of azimuthal velocity anisotropy \betat\ of the atomic gas  in M33 at different angular resolutions. 
  Solid lines show the anisotropy profiles obtained assuming \sz=\slosmean\ and a null thermal component. 
  Short-dashed lines show the $\pm$1 rms errors; a dotted line is an illustration of result  choosing another vertical dispersion, $\sigma_z = 4$ \kms\ (70 pc resolution only); and a blue long-dashed 
  line is the velocity anisotropy profile expected from the epicycle theory (Sect.~\ref{sec:discussion}).  Shaded areas are as in Fig.~\ref{fig:testisotropy}.}
\label{fig:anishi}
 \end{figure} 
 
  \begin{figure*}[t!]
   \centering
   \includegraphics[width=\textwidth]{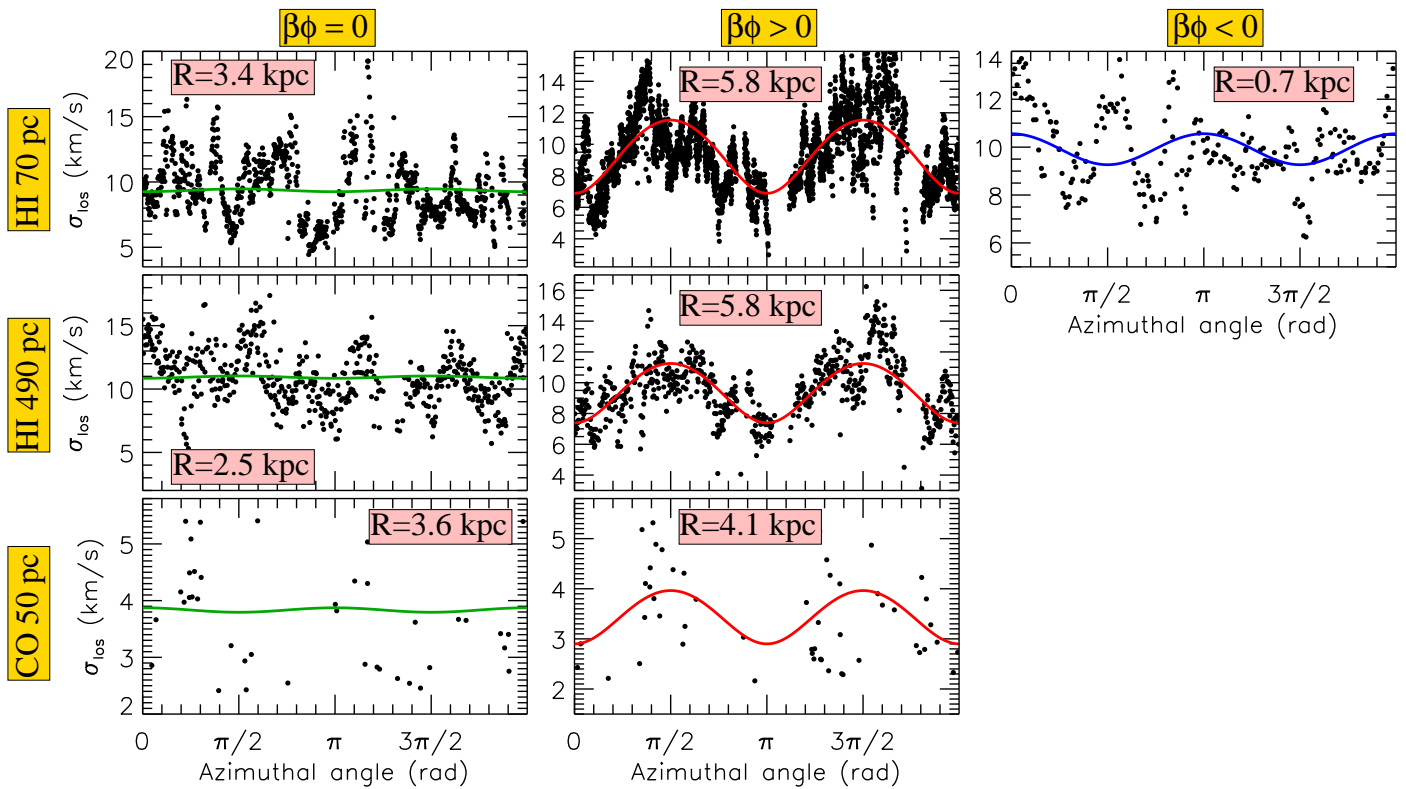} 
   \caption{Same as Fig.~\ref{fig:plotazimisoasym}, but using \hi\ data from VLA  (70 pc resolution),  DRAO (490 pc resolution), and CO data 
   from IRAM 30m antenna (50 pc resolution) to show  results of the anisotropic velocity dispersion model (Sect.~\ref{sec:anisomodel}).
   The columns illustrate various cases of velocity anisotropy parameter found by the best-fit model: $\beta_\phi \sim 0$ (for isotropy, left),  $0 < \beta_\phi$ (for  radial bias, middle),  
   $\beta_\phi < 0$ (for supposedly tangential bias, right). 
   For each resolution, the selected radii are among the best positions illustrating each case of anisotropy parameter, following Fig.~\ref{fig:anishi}. 
   Solid lines represent the best-fit dispersion model at the considered radii, assuming \sz=\slosmean.   }
   \label{fig:plotazim}
   \end{figure*}

   The resulting anisotropy parameter in the disk mid-plane is shown in Fig.~\ref{fig:anishi} for the three resolutions of the \hi\ data, again for \sz=\slosmean\ (solid lines). 
   The result for another value, $\sigma_z=4$ \kms, for the 70 pc resolution data is also given (dotted line). It shows
   that the impact  on \betat\ of the choice of $\sigma_z$ is negligible as differences in anisotropy parameter $ \lesssim 0.15$ are  observed. 
  A similar finding with stellar velocity anisotropy was presented in \citet{che18}. 
   
  The anisotropy parameters of the VLA 70 and 100 pc resolution \hi\ data match perfectly. 
  In the inner $R=3.5$ kpc, \betat\  is highly variable,  sometimes corresponding to isotropic-to-radial motions ($\lesssim 0.25 $), sometimes more radially biased ($\sim 0.4$), or showing 
    dips down to $-0.4$ ($R \sim 0.7$ and 2.2 kpc). 
    Beyond $R = 3.5$ kpc, \betat\ increases steadily reaching $\sim 0.8$, showing that orbits become more radial at large radii.
   
  The radially oriented orbits at large radius are also observed in the 490 pc resolution DRAO data (bottom panel of Fig.~\ref{fig:anishi}).
  The variations occurring on small angular scales are lost, however, because of the lower resolution. 
  The dip of \betat\ at $R \sim 0.7$ kpc is not observed, while the second dip seems to be detected at $\sim 2.5$ kpc, although at a lower (absolute) amplitude than 
  at higher resolution, making the velocity ellipsoid more consistent with isotropy.   
  Given the coincidence of the dips within or close to the regions where $\sigma_2$ was measured lower,  we interpret such 
  negative values as a failure of the anisotropic velocity model at these locations.

   The effects of isotropic and radial orbits on \slos\ are shown in   azimuth-dispersion diagrams of Fig.~\ref{fig:plotazim} (solid lines), as  extracted from the   70 pc   and 490 pc resolution \hi\ observations. 
   The widths of the radial rings were set to 0.1-0.2 kpc for the 70   pc resolution and 0.3-0.5 kpc for the 490 pc resolution.
    The left column illustrates locations where the anisotropic models found isotropy, the middle column   
   is for gas orbits that are found more radial, and the right column illustrates one of the locations of a hypothetical tangentially biased orbit in the VLA data.
   While  the model dispersion clearly varies with $\phi$   on a large angular scale, 
    it does not account for the variations seen on smaller scales.

\begin{figure}[t!]
   \centering
  \includegraphics[width=\columnwidth]{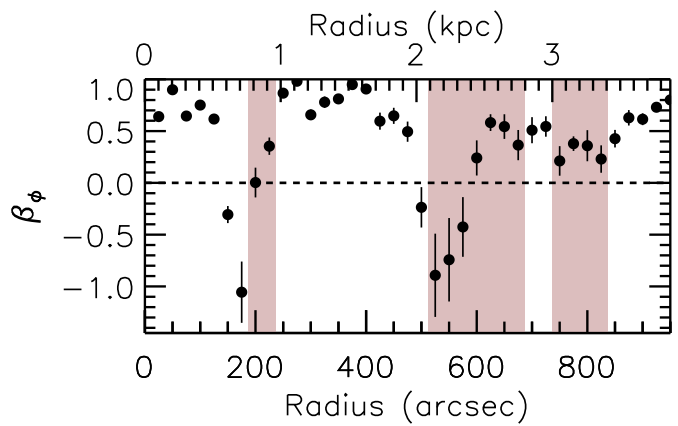} 
  \caption{Profile of azimuthal anisotropy \betat\ of the velocity dispersion of the atomic gas in M33 obtained using the 100 pc resolution VLA data by assuming \sz=\slosmean\  
  and \hi\ as a warm neutral medium (thermal component of 6 \kms). Shaded areas are as in Fig.~\ref{fig:testisotropy}.}
\label{fig:anishiwarm}
 \end{figure}

  As for the case of \hi\ seen as a warm neutral medium ($\sigma_T= 6$ \kms), \betat\ 
   is shown in Fig.~\ref{fig:anishiwarm}, again assuming \sz$=$\slosmean. The anisotropy could not  
  be derived for the outer disk because there is  little room left for both planar components. 
  This shows that at least some of the \hi\ is cool in the outer region within the anisotropy assumption. The remaining velocity anisotropy is
  stronger than in the case $\sigma_T= 0$ \kms, corresponding to even more radially biased gas orbits.

  \subsection{Results for the molecular gas}
 \label{sec:resultmol}
  
\begin{figure}[t!]
   \centering
  \includegraphics[width=\columnwidth]{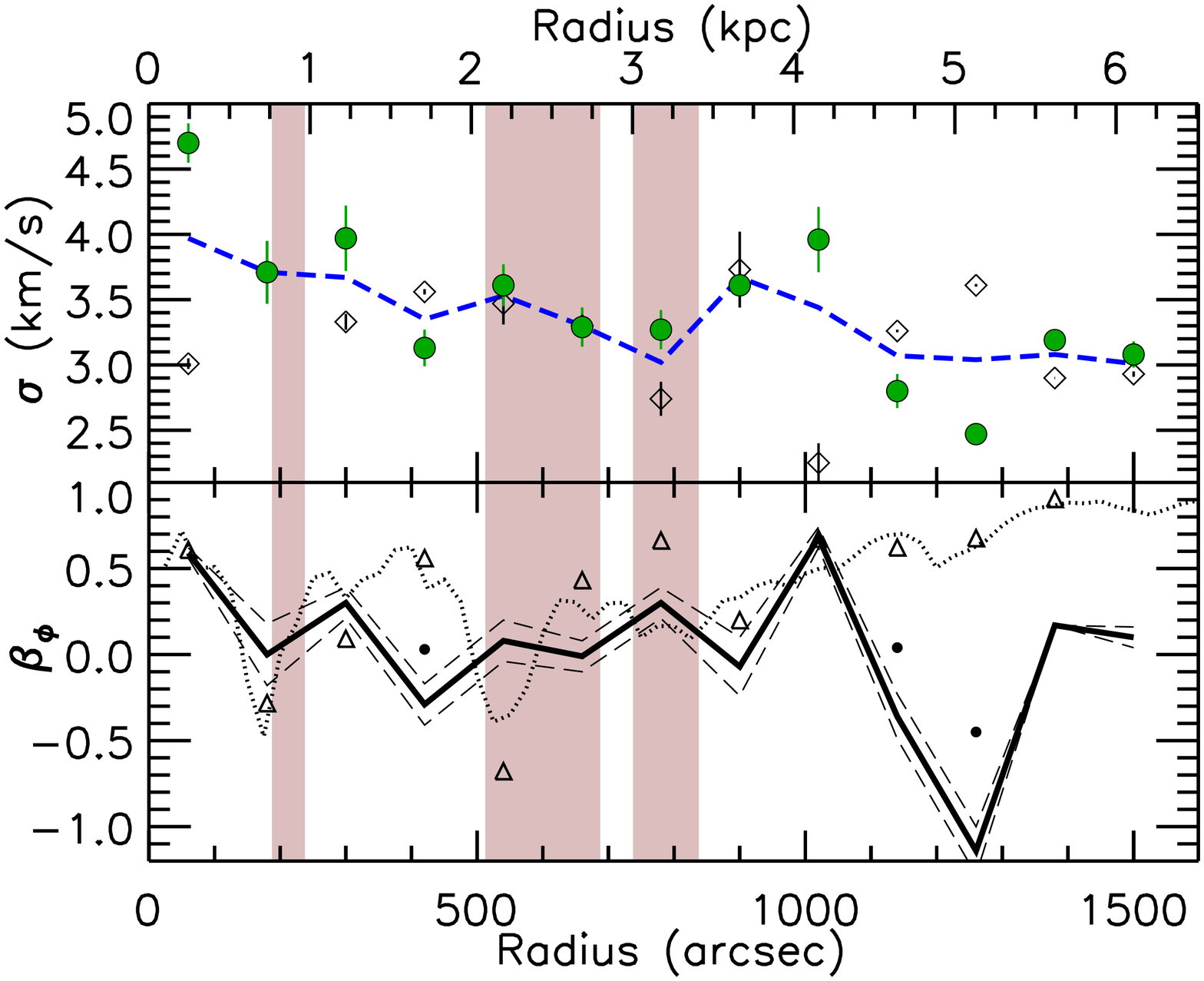}
  \caption{Results for the molecular gas in Messier 33. Top: Profiles of \sr\  and \st\ of CO gas, obtained assuming \sz=\slosmean\ (dashed blue line, corrected for instrumental dispersion) and a null thermal component. 
  Filled circles are for \sr\ and open diamonds are for \spp. 
  {Bottom:} Profiles of azimuthal velocity anisotropy \betat\ of CO gas. 
  The solid line shows the anisotropy obtained assuming \sz=\slosmean\ and a null thermal component, and  dashed lines show the $\pm$1 rms errors.
  For comparison, the anisotropy of the atomic gas as derived from all pixels in the 100 pc resolution velocity dispersion map is shown as a dotted line, 
  and that derived using only the pixels at the positions of the molecular clouds as open triangles. 
  Filled circles  are values derived by masking a few deviant dispersions of CO clouds ($\sigma_{\rm los} > 5$ \kms). Shaded areas are as in Fig.~\ref{fig:testisotropy}.}
 \label{fig:anismol}
 \end{figure}

 Results of the two-parameter model for the molecular gas, derived from the CO data, are shown in Fig.~\ref{fig:anismol}, obtained assuming \sz=\slosmean. 
  Examples of azimuth-dispersion diagrams with the \slos\ models are shown in Fig.~\ref{fig:plotazim} {(ring of 0.35 kpc in width)}.   
  Similarly to the atomic gas, the variation of the molecular gas velocity anisotropy as a function of $\sigma_z$ is negligible within the spanned range of \sz. 
  The  velocity anisotropy parameter is highly scattered ($\langle \beta_\phi \rangle \sim 0$, on average, throughout the disk).

  We inspected the origin of the significant \betat\ troughs at $R=1.7, 4.6$ and 5.1 kpc
  and found they are caused by the presence of a few deviant observed dispersions, namely $\sigma_{\rm los} \sim 6.2 $ \kms\ along $\phi = 0$ ($R=1.7$ kpc), 
  $\sigma_{\rm los} \sim 7.2 $ \kms\ along $\phi = 0$ ($R=4.6$ kpc), 
  or $\sigma_{\rm los} \sim 6.4 $ \kms\ along $\phi = \pi$ ($R=5.1$ kpc), while at these radii the other points are almost exclusively below 4.5 \kms.  
  The masking of such outlying points made the anisotropy parameter closer to 0 (filled circles
  in Fig~\ref{fig:anismol}). A close inspection of \slos\ at $R=5.1$ kpc shows no clear sine pattern that could yield $\beta_\phi < 0$. 
  Therefore, as for the \hi, there is no  evidence of a  strong tangential bias of velocity anisotropy in the molecular gas in M33. 
    
  The CO velocity anisotropy parameter is roughly in agreement with that of the atomic gas (dotted line) in the inner region, with the exception of $R=1.7$ kpc. 
  Interestingly, at larger radius where the \hi\ velocity dispersion has become strongly anisotropic, the orbital structure of CO and \hi\  in the mid-plane differ fundamentally.  
    To verify whether these differences may be artifacts, we measured the velocity anisotropy of the atomic gas in a comparable way to that of  the CO gas. This was done
  by using only the \hi\ velocity dispersions at the locations of the discrete molecular clouds, 
  instead of the whole \hi\ dispersion map.
 The result shown as open triangles in Fig.~\ref{fig:anismol} indicates that this \hi\ anisotropy profile perfectly agrees with  
 the profile inferred using the whole velocity dispersion field, including the outer regions with stronger \hi\ radial bias. 
 This suggests that the sparser distribution of points in the disk for the molecular gas than for the atomic gas 
 is not the cause of the molecular-atomic difference. 
 
 In summary, we find no compelling evidence for a velocity ellipsoid of the  molecular clouds being  
   aligned systematically towards (or perpendicular to) the direction of the galactic centre of M33. This result may indicate  that the dynamics of clouds is locally 
   dominated by the cloud gravitational
   potential. It also highlights the need for velocity dispersion maps of molecular gas in galaxies rather than cloud-based measurements 
   to make the comparison with the \hi\ gas more appropriate.
 
  \section{Discussion}
   \label{sec:discussion}
   
This work is the first to our knowledge that examines the effects of the collisionless medium hypothesis for gas on the structure of velocity dispersions, and
the implied azimuthal-to-radial axis ratio of the velocity ellipsoid in galactic disks. Therefore, no fully appropriate comparisons with other observational studies of the gas component are available. 

Comparisons can be made with the stellar collisionless kinematic tracer, however. 
Observations of radially biased stellar random motions is not rare among nearby spiral galaxies.
Taking the example of our Galaxy, the velocity anisotropy of stars in the  disk of the Milky Way derived by \citet{che18} using stellar dispersions 
from the Second Gaia Data Release published in \citet{kat18} shows quite similar values to the atomic gas of M33 inside  $R = 4$ kpc,
as well as the increase towards large radii. A possible mechanism for the origin of stellar anisotropic orbits in the Galaxy may be radial migration induced by the dynamics of 
spiral arms \citep{ros12, gra14}. The similarity of the velocity anisotropy of \hi\ gas in M33 to that of the stars in the Milky Way may provide clues for the interpretation of the results presented here. 
There are clear spiral-like structures at large radius in M33 (Fig.~\ref{fig:imam33}), 
with location and shape correlated with the bisymmetry in the velocity dispersion.
Although non-axisymmetric perturbations are not necessary to have an anisotropic velocity dispersion ellipsoid, we can speculate that 
the   spiral arms in M33 could enhance the velocity anisotropy parameter in the outer disk regions if part of the \hi\ gas in M33 behaves like a collisionless medium.
We also note  that disks of similar stellar mass to that of M33 show $-0.1 \lesssim \beta_\phi \lesssim 0.2$ \citep{che18}, 
thus stellar orbits in these galaxies are  more isotropic. 

The analysis also shows that  it is not possible to strongly constrain \sz\  given the small scatter of \betat\ as a function of $\sigma_z$. 
It is nevertheless worth mentioning that under realistic hypotheses on \sz, the  range of  $\sigma_z/\sigma_R$ found for gas in M33
is consistent with the value of $\sim 0.65$ found  for late-type stellar disks by \citet{pin18}. 

Comparisons can be made with numerical simulations as well.
Hydrodynamical modeling of gas in simulated spiral galaxies shows velocity dispersion components that are anisotropic \citep{bot03,age09}. 
In these numerical models, the tangential component is smaller than the radial dispersion. This is   
in agreement with most of our measurements. 
The simulations of \citet{age09} are interesting for our study because they are supposed to simulate a disk with similar physical properties to M33. 
They showed that the planar dispersion, \splan, given by 
the root mean squared value of \sr\ and \spp, is twice larger than \sz. 
   If we restrict the comparison to the radial range 
  $4-7.5$ kpc where M33 shows a more significant   anisotropy parameter in the framework of an axisymmetric and anisotropic velocity model, 
   the simulations of  \citet{age09}  show  \sz\ within $3.5-7$ \kms\ once the spiral-like features are well defined in the simulated density map.
 Our models show that  \splan$\sim $2\sz\ in M33 for   \sz$\sim 6$  \kms. Therefore, comparable planar and vertical dispersions are found in both the observations and simulations.
 More broadly, within  the range of  vertical dispersions that has been investigated here, we find \splan\ from  $\sim 10$ \kms\ ($\sigma_z$=10 \kms) to $\sim 13$ \kms\ ($\sigma_z$=4 \kms),
 hence \splan/\sz\ from $\sim 0.9-1$  to  $\sim 3-3.2$, respectively.   
 The vertical motion is thus the main driver of the ratio. For \sz=\slosmean\ ($\sim8$ \kms), \splan/\sz $\sim1.4-1.5$, which is  close to the value
 expected for an isotropic velocity ellipsoid ($\sqrt{2}$). 

\citet{age09} then found that  \fract\ is roughly consistent with the expectation of the epicyclic approximation (EA)  
which considers collisionless orbits that only slightly deviate from circularity. The EA stipulates that \fract\ is related to the slope of the circular velocity  $v_c$ 
     \citep[e.g.,][]{bin08}:   
   \begin{equation}
    \left(\sigma_\phi/\sigma_R\right)_{EA}^2  = \frac{1}{2} \left( 1+ \dfrac{\mathrm{d} \ln v_c}{\mathrm{d} \ln R}\right) \, . 
     \label{eq:eqapex}
    \end{equation}

 The epicycle anisotropy of M33, $\beta_{EA} = 1 - \left( \sigma_\phi/\sigma_R \right)^2_{EA}$ is shown as a dashed blue line in Fig.~\ref{fig:anishi}. It was derived using the model of the \hi\ rotation curve 
 of M33 by \citet{koc18} as a proxy for the circular velocity. The assumption {that the circular velocity can be approximated by the tangential velocity} is reasonable for gas,  
 except maybe in lower mass disks \citep{dal10}. The value of  \betaea\ shows very little variation as a function of radius.
 Interestingly, the inner $R=4$ kpc of M33 is the range of radii where the anisotropy is mild (Fig.~\ref{fig:anishi}) and differs by $\lesssim 0.25$ from that derived 
from the epicyclic approximation ({except for} the two outlying dips at $R \sim 0.7$ and $2.1$ kpc). 
The velocity anisotropy of gas is systematically larger than expected from the epicyclic approximation beyond $R \sim 5$ kpc, however. 
Equation~\ref{eq:eqapex} cannot allow the observed strongly radial \hi\ orbits because the rotation curve of M33  
is barely rising at these radii ($\beta_{EA} \sim 0.4$). We would need a decreasing rotation curve to get $\beta_{EA} > 0.5$ compatible with the observed velocity anisotropy in the plane. 
Our results thus do not fully agree with those found with the numerical simulations of \citet{age09} in the disk regions of stronger radial bias. 
On the other hand, if the ellipsoid of velocity is isotropic, then the epicyclic approximation is   violated as well because 
  a linearly rising velocity curve with radius is strictly required for an EA--isotropy agreement. This indicates a failure of the epicycle approximation of orbits for the gas component. 
Interestingly, this discrepancy is reminiscent of the result found in \citet{che18} for stellar disks where  the diversity of stellar orbits could not be reproduced by the theory. 
  
   Is the \hi\ velocity dispersion ellipsoid of Messier 33 anisotropic or isotropic? Choosing a side for this question is not an easy task as 
    the two models both explain the major asymmetry.  
    The Fourier analysis has the flexibility to probe a large number of modes, hence isotropic and asymmetric models are unsurprisingly more accurate in modeling \slos.
     An intriguing result of this work is the observation that the orientation of kinematic perturbations are often aligned with the principal disk axes (modulo half and full periods)
     as if a projection effect affected significantly the velocity dispersion map of M33. 
     Such  coincidences may be fortuitous in M33 given that asymmetries in the gas density often appear coincident with the dispersion asymmetries 
     (Fig.~\ref{fig:imam33}).     
       This also suggests that further analyses of phases of asymmetric modes inside velocity dispersion maps are promising
   to assess the nature of gaseous velocity ellipsoids. If the phase angles   were found  systematically near the
   principal axes, at the positions of supposedly stronger velocity anisotropy, then it would rule out isotropic velocity ellipsoids 
   since that occurrence should occur only rarely for a random distribution of phase angles of perturbations in the isotropic scenario. 
   Data from deep surveys like  THINGS \citep{wal08}, HALOGAS \citep{hea11},  and LittleTHINGS \citep{hun12} for the neutral atomic gas, or PHANGS-ALMA \citep{sun18} for the molecular gas 
   will be helpful to study that problem.

     \section{Summary}
   \label{sec:summary}
Messier 33 has a  non-axisymmetric distribution of observed random motions of \hi\ gas. 
 There is a prominent pattern that makes the \hi\ velocity dispersion   weaker near the major axis and stronger near the minor axis of the galaxy.  
 The velocity dispersion of the $R >4$ kpc disk  can  locally be larger by up to 60\%\ than the azimuthally averaged value. 
 Hypotheses allying axisymmetry and isotropy are ruled out to explain the variations of the velocity dispersion.
 
 Among the models presented in this study a Fourier transform has shown that bisymmetric random motions having 
 an amplitude of up to 2 \kms\ (25\%\ of the axisymmetric value) must be invoked to explain the discrepancy while maintaining isotropy. 
 It dominates the harmonic asymmetries in the random motions, and first-, third-, and fourth-order motions were found mostly weaker.  
   The phase angles of the asymmetries are often seen close to the principal axes of the \hi\ disk, and particularly the bisymmetry, which is aligned with the minor axis.   
   The asymmetries coincide well with the non-axisymmetric spiral-like distribution of \hi\ gas in M33.
   
   Another model was to consider that the velocity dispersion ellipsoid is axisymmetric but anisotropic, acting as if part of gas behaved like a collisionless medium. That led us to constrain   the 
 radial and tangential components of the ellipsoid (\sr\ and \spp) at fixed vertical dispersion \sz, and from which the azimuthal velocity anisotropy parameter $\beta_\phi=1-(\sigma_\phi/\sigma_R)^2$ 
 could be measured. In the framework of this axisymmetric model, \betat\ is mostly positive and maximum at $R\sim 6$ kpc, indicating  orbits of the atomic gas  that are strongly radial. 
 The perturbed dynamics in the spiral-like structure could be responsible for the velocity anisotropy.  
 It was also found that while anisotropic velocity dispersions could be measured when the 
 \hi\ gas is treated like a CNM, it is not the case in the outer regions of stronger velocity anisotropy 
 when the gas is seen as a WNM. A high thermal component does not leaves as  much room for the planar motions.
 
 As for the CO gas traced by 
 a collection of a few hundreds of molecular clouds, $\beta_\phi$ is highly scattered and did not allow us to draw firm conclusions about the shape of CO cloud orbits. 
 Although the comparison with \hi\ gas remains limited because cloud-based dispersions were used in this analysis, unlike \hi\ dispersions this result is not surprising 
 if  velocity dispersions of molecular clouds are driven by local cloud dynamics. 
 
These results were found  only marginally dependent on the assumptions made for \sz\ (chosen as representative of values observed in face-on nearby disks). 
  In future works, we will pursue the analysis of the properties of asymmetries in gas velocity dispersions by means of a larger sample of disk galaxies via further sensitive \hi\ and CO measurements,  
 and investigate to what extent anisotropic velocity ellipsoids  can still explain the asymmetric gas random motions.

 \begin{acknowledgements}
We are very grateful to an anonymous referee for insightful propositions which improved the analysis and the content of the article. 
This research was supported by the Comit\'e Mixto ESO-Chile and the DGI at University of Antofagasta, and J. Braine by the MINEDUC-UA project code ANT 1755. 
\end{acknowledgements}

\bibliographystyle{aa} 
\bibliography{ms-m33}

\end{document}